\documentclass[journal]{IEEEtran}
\ifCLASSINFOpdf

\else

\fi

\usepackage{amsmath,amsfonts}
\usepackage{amsthm}
\usepackage{algorithmic}
\usepackage{multirow}
\usepackage{booktabs}
\usepackage[ruled,linesnumbered]{algorithm2e}
\usepackage{array}
\usepackage[caption=false,font=normalsize,labelfont=sf,textfont=sf]{subfig}
\usepackage{textcomp}
\usepackage{stfloats}
\usepackage{url}
\usepackage{verbatim}
\usepackage{graphicx}
\usepackage[colorlinks,citecolor=blue,linkcolor=blue,urlcolor=blue]{hyperref}
\usepackage{cite}
\usepackage{cancel}
\usepackage{indentfirst}
\usepackage{cite}
\usepackage{color}
\usepackage{mdframed}
 
\usepackage{graphicx}
\usepackage{caption}

\begin{document}
	
	\title{An Fluid Antenna Array-Enabled DOA Estimation Method: End-Fire Effect Suppression}
	
	\author{Jiaji Ren, Ye Tian, Baiyang Liu, Tuo Wu, Wei Liu, \\ Kai-Kit Wong, \emph{Fellow}, \emph{IEEE}, Kin-Fai Tong, \emph{Fellow, IEEE}, \\ Kwai-Man Luk, \IEEEmembership{Life Fellow,~IEEE}
		\thanks{ \emph{(Corresponding author: Ye Tian and Kin-Fai Tong.)} } 
		
		\thanks{ This work was partially supported by the Zhejiang Provincial Natural Science Foundation of China under Grant LY23F010004, and the Natural Science Foundation of Ningbo Municipality under Grant 2024J232. The research work of K. F. Tong and Baiyang Liu was funded by the Hong Kong Metropolitan University, Staff Research Startup Fund: FRSF/2024/03. The work of Tuo Wu was funded by Hong Kong Research Grants Council under the Area of Excellence Scheme under Grant AoE/E-101/23-N. Y. Tian and J. Ren are with the Faculty of Electrical Engineering and Computer Science, Ningbo University, Ningbo 315211, China (E-mail: $\rm tianfield@126.com; 2311100053@nbu.edu.cn$). B. Liu and  K. F. Tong are with the School of Science and Technology, Hong Kong Metropolitan University, Hong Kong SAR, China. (E-mail: $\rm \{byliu,ktong\}@hkmu.edu.hk$). 
			T. Wu and K.-M. Luk are with the State Key Laboratory of Terahertz and Millimeter Waves, Department of Electronic Engineering, City University of Hong Kong, Hong Kong. (E-mail: $\rm \{tuowu2,eekmluk\}@cityu.edu.hk$).
			W. Liu is with the Department of Electrical and Electronic Engineering, Hong Kong Polytechnic University, Kowloon, Hong Kong, China (E-mail: $\rm wliu.eee@gmail.com$). 
			K.-K. Wong is with the Department of Electronic and Electrical Engineering, University College London, WC1E 6BT London, U.K., and also with the Yonsei Frontier Laboratory and the School of Integrated Technology, Yonsei University, Seoul 03722, South Korea (E-mail: $\rm kai$-$\rm kit.wong@ucl.ac.uk$).	  
		}

	}

	\markboth{IEEE TRANSACTIONS ON VEHICULAR TECHNOLOGY,~Vol.~xx, No.~x, NOVEMBER~2025}%
	{Shell \MakeLowercase{\textit{et al.}}: Bare Demo of IEEEtran.cls for IEEE Journals}
	%
	
	\maketitle
	
	\begin{abstract}
		Direction of Arrival (DOA) estimation serves as a critical sensing technology poised to play a vital role in future intelligent and ubiquitous communication systems. Despite the development of numerous mature super-resolution algorithms, the inherent end-fire effect problem in fixed antenna arrays remains inadequately addressed. This work proposed a novel array architecture composed of fluid antennas. By exploiting the spatial reconfigurability of their positions to equivalently modulate the array steering vector and integrating it with the classical MUSIC algorithm, this approach achieved high-precision DOA estimation. Simulation results demonstrated that the proposed method delivers outstanding estimation performance even in highly challenging end-fire regions.
	\end{abstract}
	
	
	\IEEEpeerreviewmaketitle
	
	\section{Introduction}
	\IEEEPARstart{D}{irection} of Arrival (DOA) estimation serves as a core enabling technology for establishing spatial awareness capabilities in six generation (6G) and beyond wireless communication systems. By resolving electromagnetic wave propagation paths with high angular precision, it provides critical support for intelligent and ubiquitous communication infrastructures. Representative applications span across diverse domains: massive multiple-input multiple-output (MIMO) beamforming in 6G terahertz communications, millimeter-wave radar perception for autonomous vehicle navigation, and distributed sensor localization within internet of things (IoT) ecosystems \cite{r1},\cite{r2},\cite{r3}. To satisfy the increasingly stringent performance requirements for DOA estimation accuracy in these emerging application scenarios, current systems must urgently overcome the fundamental physical constraints inherent in conventional antenna arrays.

	Conventional fixed antenna arrays exhibit an inherent end-fire effect that imposes fundamental physical constraints on DOA estimation. When incident angles approach the end-fire region (e.g., $[80^{\circ}, 90^{\circ}]$), the sensitivity of the steering vector to angular variations decreases significantly, leading to a sharp deterioration in spatial discriminability \cite{r4}. Consequently, most DOA algorithms fail to guarantee effective source resolution or reliable estimation accuracy under such conditions. 
	
	Numerous mature DOA estimation techniques have been developed for fixed arrays. Among these, the Multiple Signal Classification (MUSIC) algorithm has garnered widespread adoption owing to its super-resolution capability and computational tractability. Nevertheless, MUSIC inherently cannot overcome the end-fire effect, resulting in a constrained effective field of view (FoV) for practical direction finding. Although various MUSIC derivatives (e.g., real-valued MUSIC \cite{r6}, root-MUSIC \cite{r7}) have subsequently emerged, these variants primarily optimize computational complexity without achieving comprehensive performance superiority over classical MUSIC in estimation accuracy. Crucially, none can transcend the physical limitations imposed by the end-fire effect.
	
	The mitigation of the end-fire effect primarily involves three categories of approaches. \textbf{First}, expanding the array aperture through either increasing the number of antennas to form large-scale arrays or designing sparse arrays to obtain large equivalent virtual apertures \cite{r8},\cite{r9}. Although large-scale arrays can mitigate the end-fire effect to some extent through enhanced steering vector characteristics, the fundamental problem persists and continues to affect DOA estimation performance, especially at extreme angles. \textbf{Second}, steering vector mapping techniques (e.g., manifold separation technique \cite{r10}) transform the physical array into a virtual uniform linear array more conducive to signal processing. While this approach circumvents the end-fire effect and maintains applicability to arbitrary physical configurations, it requires customized mapping matrices for specific array geometries, and achieving precise mapping for complex scenarios remains computationally challenging. \textbf{Third}, co-design approaches that jointly optimize array configuration and algorithms to eliminate the end-fire effect in specific scenarios (e.g., \cite{r11}, \cite{r12}). However, such methods suffer from limited generalizability due to their dependence on fixed array geometries and often demand significant additional hardware support. In summary, existing solutions exhibit inherent limitations in terms of scalability, computational complexity, or generalizability.

	In recent years, fluid antenna systems (FAS) \cite{r13,r17,r18,r19}, representing a revolutionary advancement in antenna technology, have emerged with flexible reconfigurability and dynamic adaptability, opening new opportunities for wireless positioning. Unlike conventional fixed-position antenna (FPA) arrays where element locations are permanently static, the broader fluid antennas (FAs) framework encompasses various technologies—including electronically switchable arrays, reconfigurable metasurfaces, and spatially distributed antenna pixels—that achieve functional flexibility without requiring physical movement of conductive fluids. By enabling dynamic control over spatial radiating elements, FAS effectively expands the spatial degrees of freedom (DoF), creating virtual apertures that surpass the physical limitations of traditional antenna arrays. This reconfigurability enables FA arrays to dynamically modulate array manifold characteristics, thereby fundamentally circumventing the end-fire effect problem.
	
	This paper proposes an innovative FA array-assisted strategy for comprehensive end-fire effect suppression. The core innovation establishes a dynamic array manifold mapping between the reconfigurable physical array and a virtual planar array through: (i) predefining virtual equivalent angles that avoid end-fire regions, (ii) dynamically adjusting spatial positions of array elements along the $y$-axis, and (iii) creating steering vector equivalence conditions for reliable DOA estimation. When the manifold equivalence condition is satisfied, the system precisely estimates preset virtual angles using the virtual array's manifold, retrieving the true DOA through the established mapping relationship. This approach fundamentally circumvents MUSIC algorithm performance degradation in end-fire regions while maintaining computational efficiency and requiring no additional hardware beyond reconfigurable FA elements.

	\section{Signal Model}
	We consider a classical signal model with $L$ incoherent narrow-band signals incident on the array. Under a two-dimensional (2-D) azimuth-elevation angle model, the incident angle of the $\ell$-th signal is denoted as $(\theta_\ell, \phi_\ell)$, where $\theta$ represents the azimuth angle and $\phi$ the elevation angle. The received data at discrete time $t$ is modeled as
	
	\begin{equation}\label{1}
		\mathbf{y}(t) = \mathbf{A}\mathbf{s}(t) + \mathbf{n}(t),\;\;t \in [0,{T} - 1],
	\end{equation}
	where $T$ denotes the number of snapshots and
	\begin{equation}\label{2}
		\mathbf{y}(t) = [y_1(t), y_2(t), \ldots ,y_N(t)]^T,
	\end{equation}
	\begin{equation}\label{3}
		\mathbf{s}(t) = [s_1(t), s_2(t), \ldots ,s_L(t)]^T,
	\end{equation}
	\begin{equation}\label{4}
		\mathbf{n}(t) = [n_1(t), n_2(t), \ldots ,n_N(t)]^T,
	\end{equation}
	where $N$ denotes the total number of elements in the array. The additive noise $\mathbf{n}(t)$ is assumed to be white Gaussian distributed and statistically independent of the source signals. The array manifold matrix $\mathbf{A}$ is formed by concatenating steering vectors column-wise as
	\begin{equation}\label{5}
		\mathbf{A} = [\mathbf{a}(\theta_1,\phi_1), \cdots, \mathbf{a}(\theta_L,\phi_L)],
	\end{equation}
	
	The steering vector $\mathbf{a}(\theta_\ell,\phi_\ell)$ is intrinsically determined by the array geometry. For a conventional uniform planar array (UPA) deployed on the $x$-$z$ plane, the steering vector admits the explicit expression as
	\begin{align}\label{6}
		\mathbf{a}_\mathrm{UPA}(\theta,\phi) &= [1, \ldots, e^{j2\pi d (n_x\cos\theta\cos\phi+n_z\sin\phi)/\lambda}, \nonumber\\
		&\quad \ldots, e^{j2\pi d((N_x-1)\cos\theta\cos\phi+(N_z-1)\sin\phi)/\lambda}]^{T},
	\end{align}
	where $\lambda$ denotes the carrier wavelength and $d=\lambda/2$ represents the inter-element spacing. $N_x$ and $N_z$ denote the number of antenna elements arranged along the $x$-axis and $z$-axis, respectively, with $N =N_x\times N_z$. 
	
	To overcome the limitations of conventional fixed arrays, we consider a fluid antenna (FA)-enabled array architecture, where each antenna element can be dynamically repositioned along the $y$-axis while maintaining constant $x$- and $z$-axis coordinates within the Cartesian coordinate system. Let $d_n$ denote the relative displacement of the $n$-th antenna from the $x$-$z$ plane. The spatial phase shift component along the $y$-axis in the steering vector is expressed as
	\begin{equation}\label{7}
		\mathbf{a}_y(\theta,\phi) = [e^{j2\pi d_1 (\sin\theta\cos\phi)/\lambda}, \ldots, e^{j2\pi d_N (\sin\theta\cos\phi)/\lambda}]^T.
	\end{equation}
	
	Consequently, the steering vector for the FA array is modeled as
	\begin{equation}\label{8}
		\mathbf{a}(\theta,\phi) = \mathbf{a}_\mathrm{UPA}(\theta,\phi) \odot \mathbf{a}_y(\theta,\phi),
	\end{equation}
	where $\odot$ denotes the Hadamard product.
	
	Compared to fixed arrays, the FA array enables reconfigurability of the steering vector through dynamic $y$-axis displacement. This unique capability specifically addresses the limitations of traditional end-fire ambiguity by adaptively adjusting the array topology to establish intentional nonlinear phase gradients in the end-fire region, thereby enhancing angular resolution and spatial discrimination in DOA estimation.
	\section{The Proposed Method}
	\subsection{MUSIC Review}
	Given the signal model in \eqref{1}, the sample covariance matrix is computed as
	\begin{equation}\label{9}
		\mathbf{R}=\frac{1}{T}\sum_{t=0}^{T-1}\mathbf{y}(t)\mathbf{y}^H(t)=\mathbf{A}\mathbf{R}_s\mathbf{A}^H+\mathbf{R}_n,
	\end{equation}
	where $\mathbf{R}_s$ and $\mathbf{R}_n$ represent the covariance matrices of the signal and noise, respectively. 
	
	Performing eigenvalue decomposition (EVD) on $\mathbf{R}$ yields
	\begin{equation}\label{10}
		\mathbf{R} =\mathbf{U}_s \mathbf{\Sigma}_s \mathbf{U}_s^{H} + \mathbf{U}_n \mathbf{\Sigma}_n \mathbf{U}_n^{H},
	\end{equation}
	where the first and second components correspond to the signal and noise subspaces, respectively. Here, $\mathbf{\Sigma}_s\in \mathbb{C}^{L\times L}$ and $\mathbf{\Sigma}_n\in \mathbb{C}^{(N-L)\times (N-L)}$ are diagonal matrices consisting of the $L$ largest eigenvalues and the remaining eigenvalues, respectively. $\mathbf{U}_s\in \mathbb{C}^{N\times L}$ and $\mathbf{U}_n \in \mathbb{C}^{N \times (N - L)}$ contain the corresponding eigenvectors.
	
	Leveraging the inherent orthogonality between the array manifold vectors and the noise subspace, the DOA estimation results $(\hat \theta_\ell,\hat \phi_\ell)$ can be obtained at the $L$ local maximum points of the spatial spectrum function 
	\begin{equation}\label{11}
		P(\theta,\phi)=\frac{1}{{\mathbf{a}}^H(\theta ,\phi ){\mathbf{U}}_{n}{\mathbf{U}}_{n}^H{\mathbf{a}}(\theta,\phi)}.
	\end{equation}
	\subsection{End-fire Effect}
	In the MUSIC algorithm, when the search angle matches the true DOA, a pronounced peak appears in the spatial spectrum. However, as the incident angle approaches the end-fire region, the sensitivity of the steering vector to angular variations decreases significantly, leading to a sharp deterioration in spatial discriminability. This phenomenon makes it difficult for the MUSIC algorithm to generate distinguishable spectral peaks near the true angle. 
	
	To illustrate this issue, consider a $10\times 10$ UPA as an example. With a fixed incident angle in the end-fire region, the steering vector correlation is defined as
	\begin{equation}\label{12}
		\eta=\frac{1}{N^2}|\mathbf{a}^H(\theta ,\phi )\mathbf{a}(\theta_\mathrm{true} ,\phi_\mathrm{true} )|^2.
	\end{equation}
	\begin{figure}[!t]
		\centering
		\begin{minipage}{0.25\textwidth} 
			\centering
			\includegraphics[width=\linewidth]{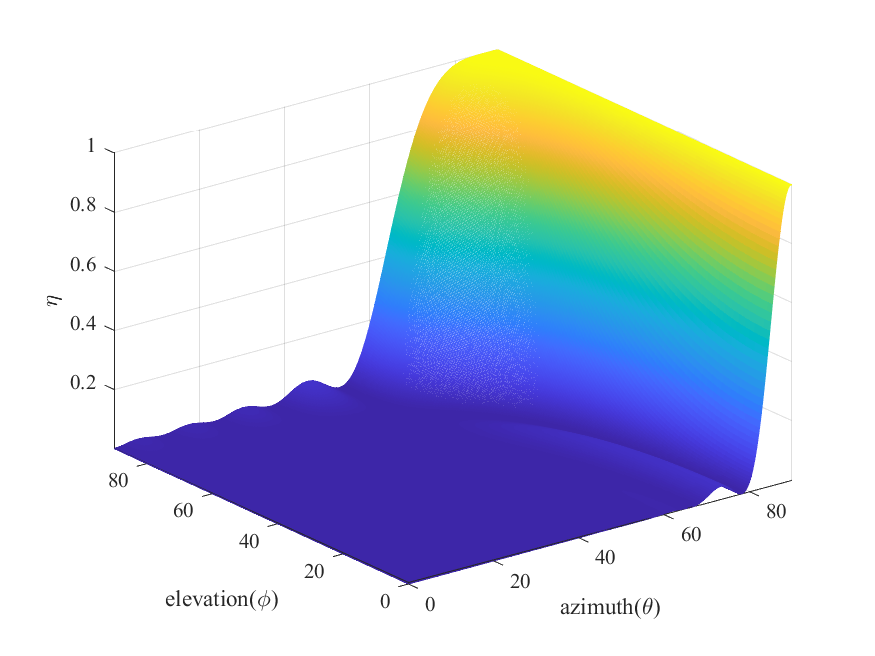}
		\end{minipage}%
		\begin{minipage}{0.25\textwidth}
			\centering
			\includegraphics[width=\linewidth]{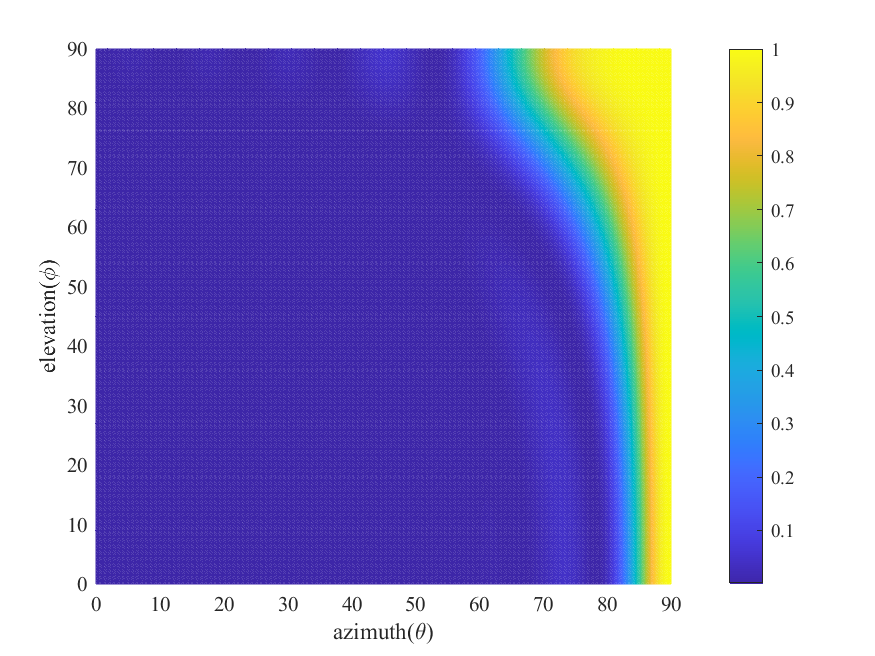}
		\end{minipage}
		
		\caption{Comparison of steering vector correlations for different angular pairs in a UPA, with $(\theta_\mathrm{true} ,\phi_\mathrm{true})$ fixed at $(86^\circ, 86^\circ)$.}
		\label{Fig.1}
	\end{figure}
	
	As shown in Fig. \ref{Fig.1}, the steering vectors constructed from angular pairs corresponding to the bright yellow regions exhibit extremely high correlation with the true steering vectors. This indicates that effective angular resolution is difficult to achieve in the vicinity of the true angle within the MUSIC spatial spectrum. Although increasing the number of antennas can partially mitigate the end-fire effect, the performance improvement achieved at higher hardware costs remains limited. Even with a $10\times10$ UPA—which itself constitutes a large-scale antenna configuration—resolution degradation in the end-fire regions remains significant.
	
	\subsection{FA-Enabled Array Equivalence and DOA Estimation}
	This subsection presents the core methodology for addressing the end-fire effect using fluid antenna arrays. Based on the steering vectors given in \eqref{6}, \eqref{7} and \eqref{8}, we extract the phase parameters and remove integer multiples of $2\pi$ to obtain
	\begin{equation}\label{13}
		\mathbf{\Phi} = \frac{\cos\theta\cos\phi}{\lambda}\mathbf{d}_x+\frac{\sin\phi}{\lambda}\mathbf{d}_z +\frac{\sin\theta\cos\phi}{\lambda}\mathbf{d}_y,
	\end{equation}
	where the position vectors are defined as
	\begin{equation}\label{14}
		\mathbf{d}_x = d[\underbrace{0,\ldots,0}_{N_z},1,\ldots,1,\ldots,(N_x-1),\ldots,(N_x-1)]^T,
	\end{equation}
	\begin{equation}\label{15}
		\mathbf{d}_z = d[\underbrace{0,1,\ldots,(N_z-1)}_{N_x},\ldots,0,1,\ldots,(N_z-1)]^T,
	\end{equation}
	\begin{equation}\label{16}
		\mathbf{d}_y = [d_1,d_2,\ldots,d_N]^T.
	\end{equation}
	Similarly, based on \eqref{6}, the phase parameters of the UPA steering vector are given by
	\begin{equation}\label{17}
		\mathbf{\Phi}_\mathrm{UPA} = \frac{\cos\theta\cos\phi}{\lambda}\mathbf{d}_x+\frac{\sin\phi}{\lambda}\mathbf{d}_z.
	\end{equation}
	
	Comparing \eqref{13} and \eqref{17}, the fluid antenna array enhances traditional architectures by introducing additional degrees of freedom through $y$-axis control. Specifically, when incident waves arrive from end-fire regions, the fixed steering vector configuration of a UPA perpetuates the end-fire effect at the physical level. In contrast, the fluid antenna array dynamically adjusts the $y$-axis positions of each element (parameter $\mathbf{d}_y$), thereby constructing a steering vector equivalent to that of a virtual incident angle. 
	
	Let the true incident angle be denoted as $(\theta_t,\phi_t)$ and the virtual incident angle as $(\theta_v,\phi_v)$. To achieve steering vector equivalence $\mathbf{a}(\theta_t,\phi_t)=\mathbf{a}_\mathrm{UPA}(\theta_v,\phi_v)$, the position adjustment parameter $\mathbf{d}_y$ must satisfy
	\begin{equation}\label{18}
		\mathbf{d}_y = \frac{\mathbf{d}_x(\cos\theta_v\cos\phi_v-\cos\theta_t\cos\phi_t)+\mathbf{d}_z(\sin\phi_v-\sin\phi_t)}{\sin\theta_t\cos\phi_t}.
	\end{equation}
	
	When condition \eqref{18} is satisfied, the FA array can be configured to be equivalent to a UPA observing a virtual incident angle $(\theta_v,\phi_v)$. This virtual incident angle can be freely chosen to avoid the end-fire region. In practical DOA estimation, the procedure involves: (i) adjusting antenna positions according to \eqref{18}, (ii) collecting signals at different positions, and (iii) performing DOA estimation using the classical MUSIC algorithm with equations \eqref{6}, \eqref{9}, \eqref{10} and \eqref{11}. If the preset virtual angle is successfully estimated, the true DOA can be retrieved through the established mapping relationship. The complete procedure is summarized in \textbf{Algorithm 1}.
	
	\begin{algorithm}[!t]
		\caption{DOA Estimation for FA Array}
		\LinesNumbered
		\KwIn{Preset virtual incident angle $(\theta_v,\phi_v)$}
		\KwOut{Estimated DOA pairs $(\hat \theta _\ell,\hat \phi _\ell)$, $\ell = 1, \ldots, L$}
		\For{$\theta\in$ search grid}{
			\For{$\phi\in$ search grid}{
				Compute $\mathbf{d}_y$ using \eqref{18} to obtain $\{\mathbf{d}_y^k\}^{K}_{k=1}$\;
			}
		}
		Configure the array according to trajectory $\{\mathbf{d}_y^k\}^{K}_{k=1}$\;
		Acquire received signals $\{\mathbf{y}_k(t)\}^{K}_{k=1}$ at each configuration\;
		\For{each $\mathbf{y}_k(t)\in\{\mathbf{y}_k(t)\}^{K}_{k=1}$}{
			Compute covariance matrix and perform EVD\;
			Construct virtual UPA steering vector using \eqref{6}\;
			Perform spectral peak search using \eqref{11} to obtain dominant peak $(\bar \theta_k, \bar \phi_k)$\;
			Compute estimation error $\epsilon_k = \sqrt{(\bar \theta _k - \theta_v)^2+(\bar \phi _k - \phi_v)^2}/\sqrt{2}$\;
		}
		Identify indices corresponding to the $L$ smallest values in $\{\epsilon_k\}_{k=1}^K$\;
		\Return{2-D DOA pairs $\{(\hat \theta _\ell,\hat \phi _\ell)\}_{\ell = 1}^L$}
	\end{algorithm}
	
	\emph{Remark 1}: Although the proposed method can theoretically handle arbitrary incident angles, it is primarily designed to address angles within the end-fire region. In practical implementation, it is recommended to first employ the conventional MUSIC algorithm for initial estimation to identify whether any angles fall into the end-fire region. If confirmed, a dynamically adjusted search grid should be designed based on this coarse estimation to reduce the computational burden of array reconfiguration operations.
	
	\emph{Remark 2}: In the algorithm, spectral peak search must be performed on the received signal at each array configuration to verify whether the preset virtual angles $(\theta_v, \phi_v)$ can be accurately estimated. If a full 2-D global search strategy were adopted, the computational overhead would be prohibitively high. It should be emphasized that only a local search within the neighborhood of the preset virtual angle is required, which can significantly reduce the computational complexity while maintaining estimation accuracy.
	\section{Numerical Simulations}
	In this section, we evaluate the DOA estimation performance of the proposed algorithm through comprehensive simulations. Unless otherwise specified, the simulation parameters are configured as follows: the array is initially deployed as a UPA with elements spaced at half-wavelength intervals in the $x$-$z$ plane. Each FA element is reconfigurable and capable of moving freely along the $y$-axis direction. A 64-element array (8 elements along the $x$-axis and 8 along the $z$-axis) is employed to receive $L=1$ narrow-band signal incident from $(86^\circ, 86^\circ)$. The incident signal is modeled as a Gaussian random process with 500 snapshots. The equivalent virtual incident angle is set to $(30^\circ, 30^\circ)$. For comparative evaluation, we consider the following baseline algorithms and performance benchmarks: subspace-based algorithms (including 2-D MUSIC \cite{r14} and ESPRIT \cite{r15}), the compressed sensing-based OMP algorithm \cite{r16}, and the Cramér-Rao Lower Bound (CRLB). The probability of successful resolution (PoSR) and the root mean square error (RMSE) of DOA estimation, obtained from 500 independent Monte Carlo trials, are employed as performance metrics. The RMSE is defined as $\mathrm{RMSE}=\frac{1}{L} \sum_{\ell=1}^{L} \sqrt{\mathbb{E}\left[((\hat{\theta}_{\ell} - \theta_{\ell})^2 + (\hat{\phi}_{\ell} - \phi_{\ell})^2)/2\right]}$.
	\begin{figure}[!t]
		\centering
		\includegraphics[width=3.6in]{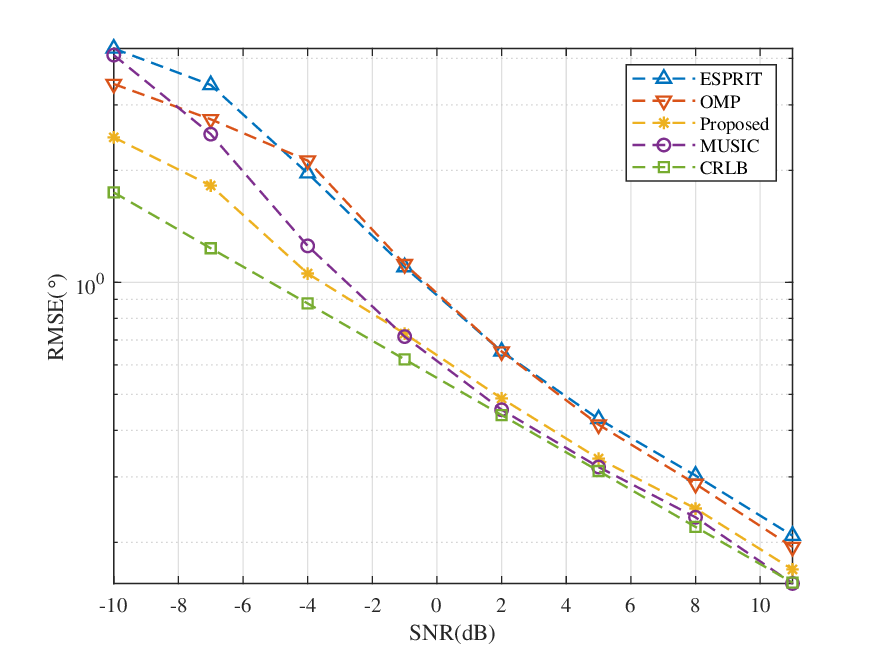}
		\caption{RMSE of DOA estimations versus SNR.}
		\label{Fig.2.}
	\end{figure}
	\begin{figure}[!t]
		\centering
		\includegraphics[width=3.6in]{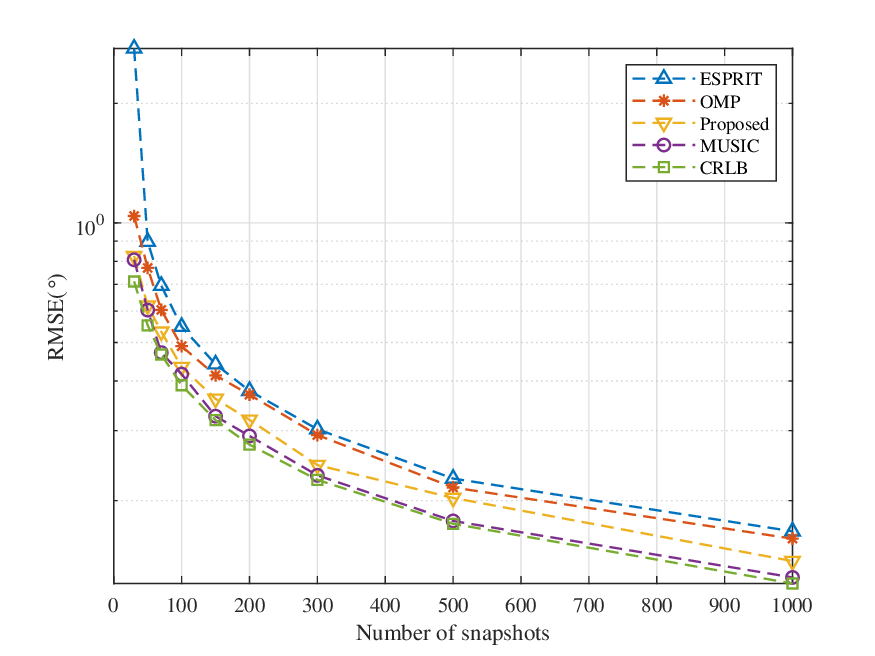}
		\caption{RMSE versus the number of snapshots $T$, with SNR = 10 dB.}
		\label{Fig.3.}
	\end{figure}
	
	\begin{figure}[!t]
		\centering
		\includegraphics[width=3.6in]{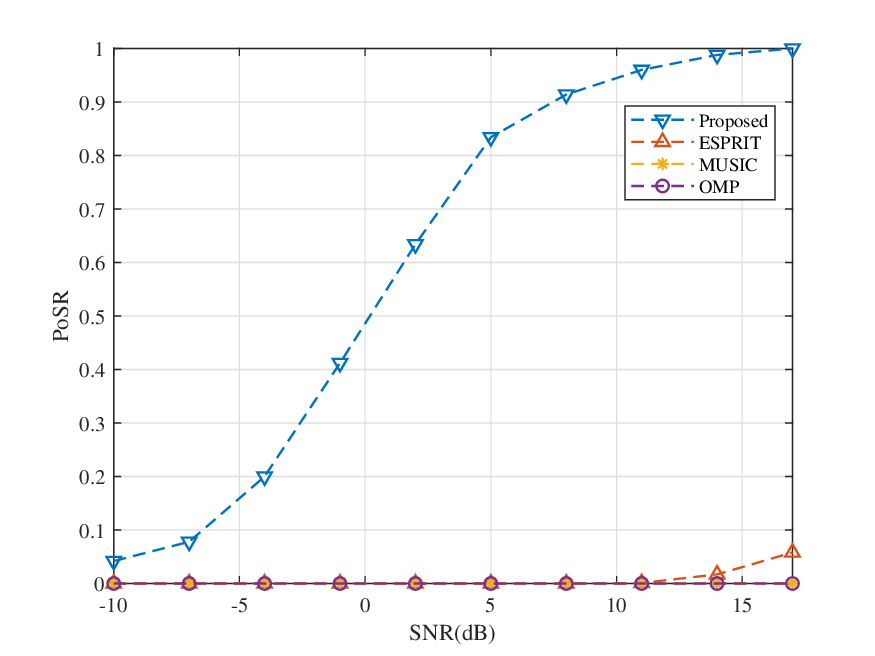}
		\caption{PoSR of DOA estimations versus SNR.}
		\label{Fig.4.}
	\end{figure}
	\begin{figure}[!t]
		\centering
		\includegraphics[width=3.6in]{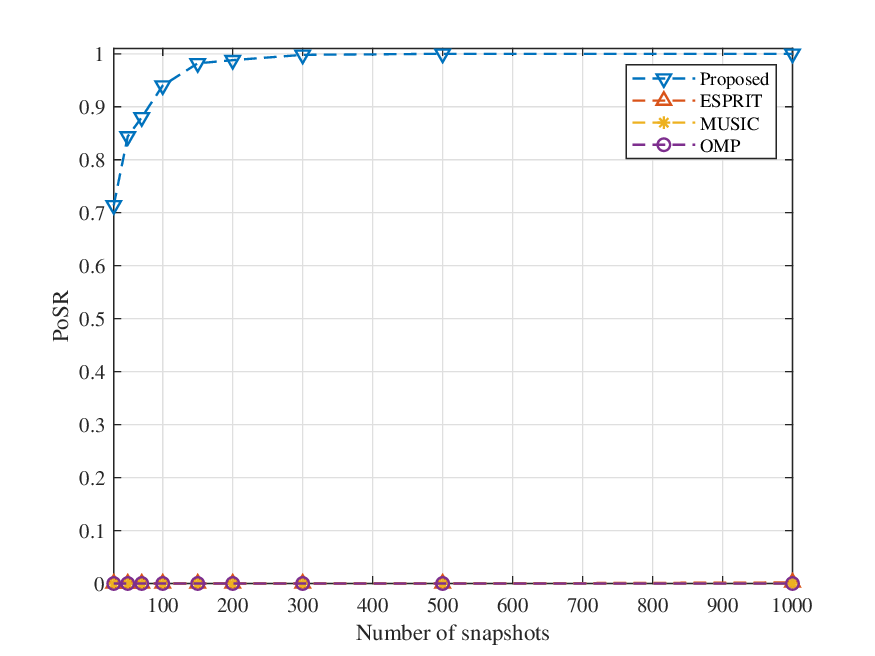}
		\caption{PoSR versus the number of snapshots $T$, with SNR = 10 dB.}
		\label{Fig.5.}
	\end{figure}
	The first simulation evaluates the DOA estimation performance of different algorithms versus signal-to-noise ratio (SNR), and the results are plotted in Fig. \ref{Fig.2.}. It is observed that the RMSE of all algorithms decreases as the SNR increases, and all achieve satisfactory angle estimation performance under high SNR conditions. The estimation accuracy of the proposed method closely matches that of the classical MUSIC algorithm and approaches the CRLB. Under conditions where conventional algorithms remain applicable, the proposed method maintains comparable estimation accuracy without noticeable performance degradation. Moreover, in the low SNR regime, the proposed algorithm demonstrates a clear performance advantage, further validating its effectiveness.
	
	The second simulation investigates the relationship between the number of snapshots and DOA estimation performance, with SNR fixed at 10 dB. As shown in Fig. \ref{Fig.3.}, the proposed algorithm achieves an angular resolution of less than $1^{\circ}$ even with a small number of snapshots. As the number of snapshots increases, the RMSE consistently decreases, eventually reaching an estimation accuracy on the order of $0.1^{\circ}$.
	
	The third set of simulations validates the ability of the proposed algorithm to resolve closely spaced sources. A second source incident from $(81^\circ, 85^\circ)$ is introduced, forming a $5^{\circ}$ angular separation with the first source. From Fig. \ref{Fig.4.}, it can be observed that even under high SNR conditions (e.g., SNR = 10 dB), the baseline algorithms almost completely fail to handle cases where two or more sources are located in the end-fire region, with PoSR approaching zero. In contrast, the multi-source resolution success rate of the proposed algorithm increases steadily with improving SNR, reaching $100\%$ estimation success at SNR = 17 dB. 
	
	Fig. \ref{Fig.5.} shows the PoSR of each algorithm when SNR is fixed at 10 dB. It can be observed that the PoSR of the baseline algorithms remains at zero and shows no improvement as the number of snapshots increases. In contrast, the PoSR of the proposed algorithm continues to rise with increasing snapshots, reaching $100\%$ estimation success when the number of snapshots reaches 300. The results indicate that only a moderate number of signal snapshots is required to achieve robust angle estimation with the reconfigurable array, significantly reducing the data volume requirements.

	\section{Conclusion}
	This paper proposed a DOA estimation method based on FA arrays to mitigate performance degradation caused by the end-fire effect. By fully leveraging the flexible reconfigurability of FAs, an equivalent mapping of the array steering vector was achieved, thereby effectively suppressing the impact of the end-fire effect. Simulation results validated the accuracy and effectiveness of the proposed method in resolving signal sources incident from the end-fire region. The proposed approach demonstrated superior performance compared to conventional algorithms, particularly in challenging scenarios with closely spaced sources in end-fire regions. Future work will focus on extending the method to more complex array geometries and investigating the practical implementation constraints of fluid antenna systems.

	

\end{document}